\documentclass[fleqn,usenatbib]{mnras}

\usepackage{newtxtext,newtxmath}

\usepackage[T1]{fontenc}
\usepackage{ae,aecompl}

\usepackage{graphicx}	
\usepackage{amsmath}	
\usepackage{amssymb}	
\usepackage{epsfig}
\usepackage{epstopdf}

\title[Cosmological Filaments]{Cosmological Filaments in the light of Excursion Set  of Saddle Points}

\author[M. Ansari Fard, S. Taamoli, S. Baghram]{
 Mohammad Ansari Fard, 
Sina Taamoli 
~and Shant Baghram, \thanks{baghram@sharif.edu}
\\
Department of Physics, Sharif University of
Technology, P.~O.~Box 11155-9161, Tehran, Iran\\\
}

\date{Accepted XXX. Received YYY; in original form ZZZ}

\pubyear{2018}

\begin{document}
\label{firstpage}
\pagerange{\pageref{firstpage}--\pageref{lastpage}}
\maketitle

\begin{abstract}
{The universe in large scales is structured as a network known as cosmic web. Filaments are one of the structural components of this web, which can be introduced as a novel probe to study the formation and evolution of structures and as a probe to study the cosmological models and to address the missing baryon problem. The aim of this work is to introduce an analytical framework to study the statistics of filaments such as number density of them and also to obtain the length-mass relation. For this objective, we model filaments as collapsed objects which have an extension in one direction, accordingly we use the  ellipsoidal collapse to study the evolution of an over-dense region via gravitational instability. 
We find that the nonlinear density of filaments in the epoch of formation is almost mass independent and is in order of $\sim 30$. By introducing {\it{filament's extended condition}} we find a fitting function for length-mass relation. For the statistics of filaments, we propose a novel framework named excursion set of saddle points. In this approach, we count the saddle points of the density field Hessian matrix, and relate it to the count of filaments. In addition, we addressed the filament in filament problem with up-crossing approximation.}
\end{abstract}

\begin{keywords}
galaxies: filaments - cosmology: theory, dark matter - methods: analytical
\end{keywords}



\section{Introduction}
It is more than two decades that the cosmological observations and simulations indicate that in the large scales the matter of the universe is distributed in the form of a web, known as cosmic web \citep{Bond:1995yt}. In this web-like structure, the filaments are one of the main components. The existence of cosmic web is a direct consequence of gravitational instability of dark matter \citep{Zeldovich:1969sb,Barrow1981}. It has been seen both in observations \citep{Shandarin:1997fc, Bharadwaj:1999jm} and simulations \citep{Springel:2006vs}. \\
The study of the physical properties of the cosmic filaments will have a great impact on our understanding about how the cosmological structures formed and evolved. On the other hand, filaments can be used as a probe for cosmological models. In this direction, recently the effect of modified gravity on the filaments is studied in \citep{Ho:2018byw}, where they used N-body simulation to find the effect of screened modified gravity on filaments shape and its geometrical properties. They found that modified gravity causes filaments to be shorter and denser with respect to the standard model of cosmology $\Lambda$CDM. As another example, \citep{Moffat:2017mak} offered that modified gravity has the ability to explain the observed mass of filaments which have been detected by gravitational weak lensing \citep{Epps:2017ddu}.\\
Another important field of investigation on filaments is related to the distribution of baryons in the universe. Baryons that inhabit in the filaments are also suggested as a solution for missing baryon problem \citep{Cen1999,Nicastro2008,Nevalainen2015}. Haider et al. show in Illustris simulation that half of the baryons are in filaments \citep{Haider:2015caa}. There are also studies in observations  \citep{Eckert:2015akf} verifying that approximately one third of the total baryons of the universe are in the filaments where they used Sunyaev Zel'dovich effect for baryon detection. 
Finally, we should mention that filaments and generally cosmic web have undeniable observational effects on late time large scale structures probes of cosmology such as weak lensing of Cosmic Microwave Background (CMB) \citep{He:2017owu}.
 The above examples show that filaments are new valuable probes for cosmological studies. \\
First of all, in order to use filaments in the cosmology, we need a precise definition for them. Up to now, several methods and techniques have been developed to account for this issue both in simulations and observations. In \citep{Hahn:2006mk}, Hessian matrix of gravitational potential, deformation tensor, used for classification of the cosmic web. The number of positive eigenvalues of deformation tensor defines the type of the corresponding cell. Halos are defined by three positive eigenvalues, and sheets/filaments correspond to one/two positive eigenvalues respectively. There is a study where instead of a positive/negative eigenvalues a threshold on eigenvalues is introduced for classification \citep{ForeroRomero:2008ig}. In \citep{Hoffman2012} the Hessian matrix of velocity field identifies the cosmic web in a similar manner to \citep{Hahn:2006mk}. Cosmic skeleton is another way to classify the cosmic web which firstly introduced in \citep{Novikov2006} and then developed in \citep{Sousbie2007}. This idea is implemented in "DisPerSE" numerical code \citep{Sousbie2011} as a cosmic web identifier. The common feature in all these three methods is that they all use saddle points of a cosmic field (gravitational potential, velocity or density field) to identify the filaments. For other approaches and techniques of filaments classifications see \citep{Libeskind2018} \\
Comparing several methods in classification of cosmic web shows that there is no agreement between these methods even from a statistical point of view. 
So theoretical models of the cosmic web could help us to face this challenge. Despite the importance of this issue, the progress in finding a theoretical model for filaments and sheets are slow in comparison to halos and voids.  (See \citep{Cooray2002} for the review on halo model and \citep{Sheth2004} for the extension of the ideas of halo identification applied to voids). {However, we should note that in \citep{Shen:2005wd}, authors used excursion set approach to find number density of filaments and also define a critical density for filament formation.
In this light "Excursion Set Theory" (EST) could be used to find the number density of filaments as the case of sheets, halos and voids.}
However, we should mention that in EST approach \citep{Shen:2005wd} the position of the formation of the filaments in density field is not considered. Accordingly, a special treatment is required to overcome this issue. Filaments tend to form in saddle point of density field (which is the basis of the cosmic skeleton method) or saddle point of velocity field or potential field. Incorporating the position of the objects in excursion set approach is feasible in the "Excursion Set of Peaks" (ESP) which have been introduced in \citep{Paranjape:2012ks} for halo formation and halo number count studies. In this work, we inspired from idea of ESP to introduce the "Excursion Set of Saddle points" (ESS) in which by using the results of ellipsoidal collapse we address the problem of the number density of filaments with ESS. We also introduce the length of filaments as an important feature of them. {We define how to find the length of filaments in ellipsoidal collapse and will show that ellipsoidal collapse find a correlation between mass and length of filaments in which the trend of this correlation is positive. However filaments are larger in simulations by a factor of 10 with respect to our model \citep{Cautun2014}. We also show that by considering the idea of "Filament's extended condition" by tuning the free parameter introduced in the model we can predict the length-mass relation of the filaments better. Intuitively, our proposed constrain would imply to have appropriate halos at the two ends of filament.\\
One important question which arises immediately is how one can use ellipsoidal collapse model to describe filaments, when they have elongated shapes already in beginning of their formation? (For more detail see \citep{Bond:1995yt}). We refer to this question as main challenge for our model. Despite the fact that considering the role of window function (spherical top-hat or Gaussian) which make density field more smooth and isotropic, our proposed model will not work
for very elongated filaments. At least we can explain this caveat by considering our model for describing those kinds of filaments which approximately were initially spherical in smoothed density field.}
The structure of this work is as follows: In Sec.(\ref{Sec:Theoretical}), we study the filaments in the context of ellipsoidal collapse and we find the length-mass relation of these structures. In Sec.(\ref{Sec:ESS}), we introduce the excursion set of saddle points as a novel idea to find the number density of filaments. In Sec.(\ref{Sec:conclusion}), we have our conclusion and future remarks. In App.(\ref{app:length}), we show the detail calculations to obtain the length of filaments and in App.({\ref{app:moving}}), we study the number density of filaments with a moving barrier condition. The results are for flat $\Lambda$CDM model with matter density of $\Omega_m=0.25$, the $H_0 =70 \, \text{km/s/Mpc}$ and $n_s = 0.95$.

\section{Ellipsoidal collapse and filament length}
\label{Sec:Theoretical}
It is almost natural to think that the ellipsoidal collapse framework is a suitable analytical model to study the collapse of dark matter structures. This model is used to find the number density of halos via the idea of the Press-Schechter (PS) \citep{Press:1973iz}. In this direction, the extensions of PS are in good agreement with simulations. Sheth-Tormen (ST) halo number density, which is firstly introduced as a fitting function and then derived form ellipsoidal collapse model, works as an example\citep{Sheth:1999mn}. In the ellipsoidal collapse model, the dark matter structures are formed from an initial homogeneous ellipsoid over-density region. However, this ellipsoid region evolves in three dimensions and collapses in an anisotropic way. The consequence of this anisotropic collapse is the formation of different types of objects in the cosmic web; sheets with one collapsed axis, filaments with two and halos with three collapsed axes. Accordingly, it will be natural to think an analogous model of halos for finding number density of filaments and sheets. {When there could be several dynamical models for describing the evolution of filaments, in which the initial region is cylindrical objects, in this work, we simply use the discussed approach (ellipsoidal collapse) to find the number density of filaments. However, we introduce a new condition for filaments formation which is the criteria of their formation at the saddle points of matter density. We also argue that constraining the Hessian matrix of potential field works better when we deal with geometrical features of the filaments such as the length of the filaments. \\}
In this direction, we solve the equation of motion of ellipsoid to find the critical density of the formation of a sheet, a filament or a halo \citep{Shen:2005wd}. We use the Zel'dovich approximation for a spherical region in Lagrangian space as an initial condition for $j$th axes of the ellipsoid, $X_j(t)$, as
\begin{equation} \label{eq:zeldovich}
X_j(t) = r_0(1 - \lambda_j D(t)),
\end{equation}
where $r_0$ is the radius of the sphere in Lagrangian space, $\lambda_j$ are eigenvalues of deformation tensor (Hessian matrix of potential) and $D(t)$ is the growth function. By definition, we set $D(t_i) = 1$ in initial time $t_i$. Eq.(\ref{eq:zeldovich}) is almost exact in the quasi-linear regime and it is valid till the time of shell crossing. It is convenient to define ellipticity $e$ and prolateness $p$ accordingly
\begin{equation} \label{eq:e-p}
e_{\phi} = \dfrac{\lambda_1 - \lambda_3}{2\delta_i},\qquad p_{\phi} = \dfrac{\lambda_1-2\lambda_2+\lambda_3}{2\delta_i},
\end{equation}
{where $\delta_i= \lambda_1 + \lambda_2 + \lambda_3$ is initial matter density contrast. So it's appropriate to find critical densities as a function of $\{e_{\phi},p_{\phi}\}$, when every thing is related to initial conditions which are encoded in these two parameters and also $\delta_i$.
\footnote{We must mention here that these ellipticity and prolateness parameters are defined due to deformation tensor and must not be confused with $e_{\delta}$ and $p_{\delta}$
which are defined from eigenvalues of the density field Hessian matrix.} Evolution of the ellipsoid's axes could be solved numerically, assuming freezing of the axes in a defined value (see App. B)}. { It has been shown that this freezing value does not affect the final results significantly \citep{BondMayers1996,Stein2018}. So when the third axis collapses, the density contrast of the overdense region will become the virial density. Accordingly, the critical density of filaments $\delta_\text{f}$ for the collapse of the second axis could be obtained approximately by}
\begin{equation} \label{eq:delta_c}
\delta_{\text{f}} = \delta_{\text{c}}(1+\alpha p_{\phi}),
\end{equation}
\begin{figure}
	\includegraphics[width=\columnwidth]{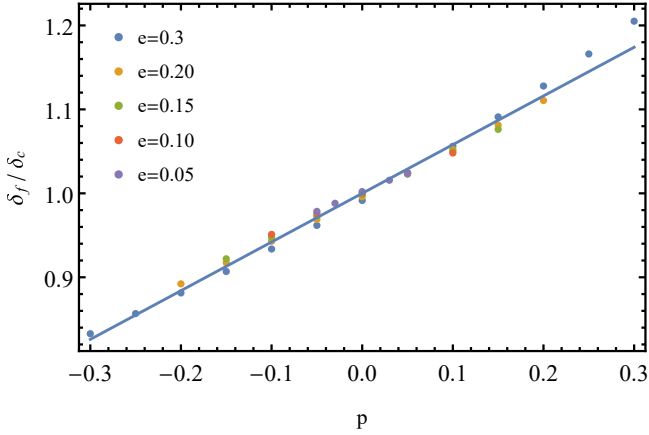}
    \caption{The critical density normalized to the spherical collapse value in the linear regime for filament formation is plotted versus $p_{\phi}$, The color index shows different vales of $e_{\phi}$.}
    \label{fig:delta_c}
\end{figure}
where $\delta_{\text{c}}\simeq 1.68$ is the spherical collapse's critical density and $\alpha$ is a constant with small dependency on the cosmology. In Fig.(\ref{fig:delta_c}), we plot the critical density of filaments normalized to $\delta_c$ in terms of $e_{\phi}$ and $p_{\phi}$ in flat Einstein de Sitter (EdS) cosmology (i.e. $\Omega_m =1$). In this figure, the best fit of Eq.(\ref{eq:delta_c}) with $\alpha \simeq 0.58$  for the data in the range of $e_{\phi}$ and $p_{\phi}$, which filaments are defined, is plotted. We should also mention that for filaments, the dependency of the critical density to ellipticity parameter $e_{\phi}$ is very weak. {In this section we will discuss that $\alpha\simeq 0$ is almost a good approximation.} \\
In initial density, each point have different $e_{\phi}$ and $p_{\phi}$ which result to different critical densities for the filament formation. However, its a good approximation, to replace $p_{\phi}$ in Eq.~({\ref{eq:delta_c}) by it's average value in the interested range \citep{Sheth:1999mn}. This approximation simplifies calculations considerably, but does not change the punch-line of the proposal. In Fig.(\ref{fig:e,p}), we plot the two dimensional space of ellipticity and prolateness. This plot will help us to categorize the large scale structures of the cosmic web. (Note that, this plot will be used for both $\{e,p\}_{\phi}$ and $\{e,p\}_{\delta}$). The ordering of eigenvalues $\lambda_1\geq \lambda_2 \geq \lambda_3$ is translated to $-e \leq p \leq e$. In Fig.(\ref{fig:e,p}), three conditions for different types of cosmic web components is considered as below\\
a) $\lambda_3 \geq 0$: condition for halos\\ b) $\lambda_2 \geq 0$ and $\lambda_3 \leq 0$: condition for filaments  \\ c)$\lambda_1 \geq 0$, $\lambda_2 \leq 0$ : condition for sheets \\ 
{d) We also consider a filament condition with extra constraint as $|\lambda_3| \geq \lambda_c$, in which $\lambda_c$ is some positive constant which for deformation tensor defined as $\lambda_{c,\phi} = \lambda_c/\delta_c$ and for Hessian tensor of density defined as  $\lambda_{c,\delta} = \lambda_c/\sqrt{s}$ (Note that $\sqrt{s}$ is the variance of perturbations). We call this new condition as filament's extended condition. Filament's extended condition implies that the derivative of density be enough large, in order to guarantee that the two peaks attached to the filament have large density. This is an approximation to impose the condition  that  filaments have two massive halos at the ending nodes. We also consider $\lambda_{c,\phi}$ and $\lambda_{c,\delta}$ as free parameters of our model which could be fixed by simulations by using  the mass-length relation of filament or/and their number density.
\begin{figure}
	\includegraphics[width=\columnwidth]{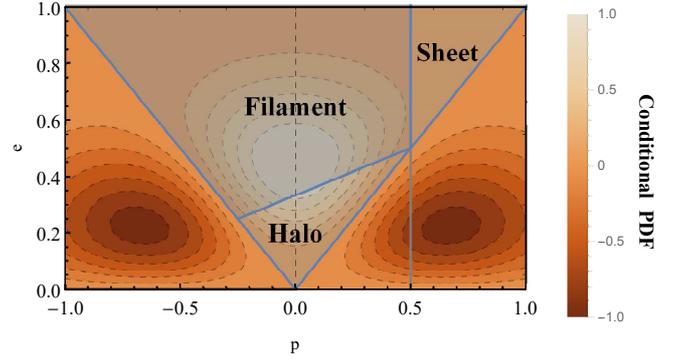}
    \caption{The two-dimensional space of ellipticity and prolaticity. The contour plots are the joint probability distribution function of $e$ and $p$. The solid lines imply the conditions on eigenvalues which can divide the two-dimensional space into the collapsed regions of filaments, halos, and sheets. }
    \label{fig:e,p}
\end{figure}
Note that in the Fig.(\ref{fig:e,p}) the contours are related to conditional joint probability distribution function (PDF) of density field's ellipticity and prolateness. (i.e. $P(e_{\phi},p_{\phi}|s, \delta_c)$  defined as\citep{Doroshkevich1970, Sheth:1999mn}.)
\begin{equation} \label{eq:distribution}
P(e_{\phi},p_{\phi}|s,\delta_c)= \dfrac{1125}{\sqrt{10\pi}}e(e^2-p^2)\left(\dfrac{\delta_c}{\sqrt{s}}\right)^5 \exp\left[ -\dfrac{5}{2} \dfrac{\delta_c^2}{s}(3e^3+p^2)\right].
\end{equation} }
The average of $p_{\phi}$ is zero when there is no condition on initial potential field (i.e. $\langle p_{\phi} \rangle_{all} = 0$). In this case the critical density is scale independent. However, considering any condition on initial potential field will lead to a scale-dependent critical density barrier. For the rest of the work we use a constant barrier for simplicity.{This assumption is equal to considering $\alpha = 0$ as we discussed earlier in this section.} The moving barrier is studied in App.(\ref{app:moving}).\\
As we have mentioned before, in the context of the ellipsoidal collapse, the formation of filaments can be considered as the collapse of the second axis of the over-dense region. Accordingly, the length of the filament is equal to the length of the third axis in time of filament formation. In Appendix (\ref{app:length}) we show that the comoving length $l(M,e_{\phi},p_{\phi})$ of filaments depends on parameters $e_{\phi}$, $p_{\phi}$ and the mass enclosed in the ellipsoidal region and it is approximately given by
\begin{equation}\label{eq:filamentlenght}
l(M,e_{\phi},p_{\phi}) = \left(\beta_W \dfrac{GM}{H_0^2 \Omega_m} \right)^{1/3} [X_v + \delta_c (e_{\phi}-p_{\phi})],
\end{equation}
where $\beta_W$ is a constant which depends to the smoothing window function. For real space top hat window function $\beta_W = 2$ while for the Gaussian window function $\beta_W = \sqrt{{8}/(9\pi)}$. Here and after we choose the Gaussian window function to be consistent with the rest of the work.
Note that $X_v$ is related to virial length of a halo $X_v^3 = {1}/{178}$. When $e_{\phi} = p_{\phi}=0$ this length tends to be the virial radius of a halo which is consistent with known previous results. In this case the collapse is spherical and all the axes collapse in the same time; accordingly the object is a spherical halo which it's radius is equal to the virial length of a halo. on the contrary, when $e_{\phi}-p_{\phi}$ is large, the initial shape of the collapsing region is filament-like (one axis is larger than other two axes). So, the final filament has larger length in comparison to the halos. The mass dependency of the length is as $M^{1/3}$  which means that the massive filaments have a larger length, consistent with our intuition. Again we could approximate Eq.(\ref{eq:filamentlenght}) by replacing $e_{\phi}$ and $p_{\phi}$ with their averages over an interested region. This adds another mass dependency to mass-length relation.\\
\begin{figure}
	\includegraphics[width=\columnwidth]{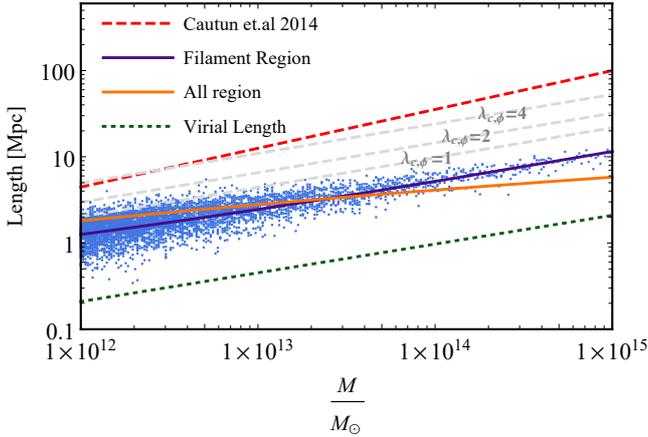}
    \caption{{The length of filaments is plotted versus mass. The solid orange line is obtained from Eq.(\ref{eq:filamentlenght}) in all regions of possible $e_{\phi}-p_{\phi}$ showed in Fig.(\ref{fig:e,p}). The blue solid line corresponds to the length of filaments (see Eq. (\ref{eq:filamentlenght})) averaged over the region in the $e_{\phi}-p_{\phi}$ plane where the filamentary condition is satisfied (see Fig.(\ref{fig:e,p})). The gray lines obtained by the  values of $|\lambda_3|/\delta_c > \lambda_{c,\phi}=1,2,4$ correspondingly (see the main text). For comparison, we plot the virial length versus mass and the prediction of Cautun et al \citep{Cautun2014}. The dotted blue data points are obtained from Monte carlo method showing the scatter about the average value of blue solid line}}
    \label{fig:length}
\end{figure}
{{ In Fig.(\ref{fig:length}), the length of filaments is plotted versus enclosed mass. The solid orange line is obtained from Eq.(\ref{eq:filamentlenght}) in the all regions of possible $e_{\phi}-p_{\phi}$ space showed in Fig.(\ref{fig:e,p}). The blue solid line corresponds to the length of filaments (see Eq. (\ref{eq:filamentlenght})) averaged over the region in the $e_{\phi}-p_{\phi}$ plane where the filamentary condition is satisfied (see Fig.(\ref{fig:e,p})). The dotted points are chosen by the Monte Carlo method, which represent the 2D scatter from average value of $e_{\phi}$ and $p_{\phi}$ with respect to distribution of the PDF which is indicated in Fig.(\ref{fig:e,p}) and Eq.(\ref{eq:distribution})). The number of points in different mass chosen by the number density of the filaments discussed in Sec.(\ref{Sec:ESS}) (see Fig.(\ref{fig:massnumber})). In this plot we also show the length to mass relation of the filaments with additional constraints $|\lambda_3| / \delta_c > \lambda_{c,\phi} = 1,2,4$. As discussed before, these new constraints are introduced to guarantee the existence of massive dark matter halo nodes in the end-points of the filaments.   We also could represent a fitting function for the average length of a filament with respect to it's mass (solid blue line of Fig.(\ref{fig:length})) }}
\begin{equation}
 \dfrac{l}{1Mps}= \Upsilon
 \left(  \dfrac{M}{M_\odot}\right) ^{\gamma},
\end{equation}
where $\Upsilon \simeq 10^{-4}$ and $\gamma \simeq 1/3 $ for flat $\Lambda$CDM cosmology.
{The above fitting function is checked for flat $\Lambda$CDM models with different matter density parameters, where we see no significant change in $\gamma$. The $\Upsilon$ parameter has small dependency to $\Omega_m h^2$, which its variation is less than $\sim 5\%$
for a $\sim 10\%$ change in $\Omega_m h^2$. The cosmology dependence of the length is not the focus of the this work, but we want to mention here that in our approach the cosmological dependence of the length and also the imprint of beyond standard models of cosmology on the length of the filaments could be investigated straightforwardly.\\
In the case of extended filament region the $\Upsilon$ is related to $\lambda_{c,\phi}$ as
\begin{equation} \label{eq:upsilon and lambda}
\Upsilon \simeq 10^{-4+0.13\lambda_{c,\phi}},
\end{equation}
where there is no meaningful change of $\gamma$ with respect to $\lambda_{c,\phi}$. The  Eq.(\ref{eq:upsilon and lambda}) present a formulae for the behavior of gray dashed lines of Fig.(\ref{fig:length}) \\
In Fig.(\ref{fig:length}) the length of filament, when averaging over all regions (no condition on potential field) is plotted with orange line. We should note that the slope of the filament's length with respect to mass, when averaging over all regions, is smaller than the pervious one. In this figure, the red dashed line is an example of length of filaments which is extracted from simulations \citep{Cautun2014}. 
Also we should note that in our approach we assume that the filaments are virialized objects, however this condition is not completely correct for filaments and they could be  extended one dimensional non virial matter clumps. In this direction, the virial length of halos is also plotted in Fig.(\ref{fig:length}) for comparison. It is interesting to note that in larger masses the average over all region line converge to virial mass.}This is a hint that we should consider the plausible region of filaments in $e_{\phi}-p_{\phi}$ plane. \\
Also at this point, by using Eq.(\ref{eq:filamentlenght}), the nonlinear virial density of filaments can be obtained as below
\begin{equation}
\delta_{NL} = \dfrac{1}{X_1 X_2 X_3} -1,
\end{equation} \label{eq:nonlineardensity}
 where $X_i$ is the anisotropic scale factor of the collapsing ellipsoid. For the collapsed filaments, two of these scale factors are equal to $X_v$ and another one is $X_v + \delta_c (e_{\phi}-p_{\phi})$.
\begin{figure}
	\includegraphics[width=\columnwidth]{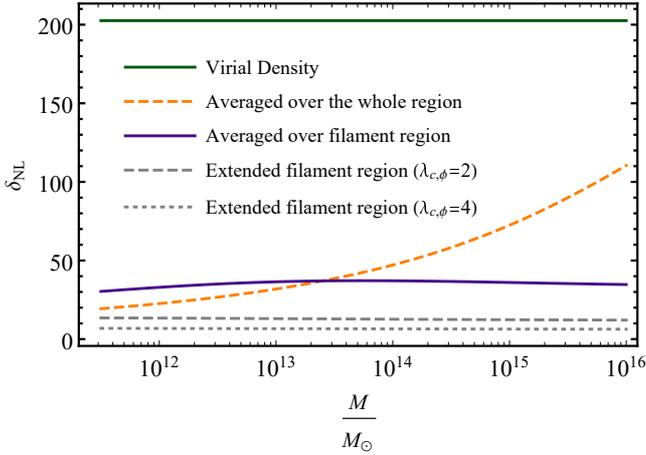}
    \caption{The non-linear density contrast is plotted versus collapsed mass. The green solid line is the density contrast for halos. The solid blue line corresponds to the filaments averaged over the filament region. The orange dashed line shows the non-linear density of filaments averaged over all possible regions.}
    \label{fig:Virial density}
\end{figure}
{In Fig.(\ref{fig:Virial density}), we plot the nonlinear over-density for halos and filaments. The blue solid line shows the case which is obtained by averaging over all filaments in the desired range of elipticity and prolateness. (This line is related to solid blue line of Fig.(\ref{fig:length})) The important point to note here is that the critical density is almost mass independent in order of $\delta_{\text{NL}}\sim 30$. This can be used as a proposed proxy to find the filaments in simulations. The dashed/dotted gray lines are for the extended filament region cases and $\lambda_{c,\phi}$ equal to $2$ and $4$ respectively. For these cases the $\delta_{\text{NL}}\sim 10, \, 5$ respectively.
The orange dashed line is the $e_{\phi}-p_{\phi}$ all regions averaging case. We see that when there is no condition on elipticity and prolateness, the non-linear density contrast is over $\sim 100$ for large masses. Once more this indicates that it is necessary to consider conditions on elipticity and prolateness.}\\
In the next section, first we briefly introduce the excursion set theory and excursion set of peaks in order to relate it to our proposal for the study of the filament's statistics.

\section{Excursion Set theory of saddle points}
\label{Sec:ESS}
Here in the first subsection, we discuss the excursion set theory in order to set the scene, then we go through  the excursion set theory of peaks and finally we propose our excursion set of saddle points as a new framework to find the number density of filaments.

\subsection{Excursion set theory}

In this subsection, we briefly review the main concepts introduced in Excursion set theory (EST) formalism  \citep{Bond:1990iw}.
In EST the number density of dark matter halos obtained by the idea of counting the trajectories which are produced in linear regime by the density contrast in terms of the scale of filtering \citep{Nikakhtar:2016bju}. The fraction of trajectories which up cross the barrier (critical density in our case) is related to the fraction of the mass which will be enclosed in dark matter halos. In order to solve the well known cloud in cloud problem, we should just count the trajectories which have their first up cross in a specific smoothing scale under consideration. In the case, if the smoothing window function is chosen to be k-space sharp filter, the process will be a Markovian. The first crossing probability with a constant barrier will be two times of the Press-Schechter \citep{Press:1973iz} prediction \citep{Bond:1990iw}
\begin{equation} \label{eq:PS}
\nu f_{\text{eps}}(\nu) = \sqrt{\dfrac{2}{\pi}} \nu e^{-\nu^2 / 2},
\end{equation}
where "eps" refer to the "extended Press-Schechter" and the height parameter is defined as $\nu={\delta_c}/{\sigma_0}$. Note that  $\delta_c$ is the critical density and $\sigma_0$ is the variance of perturbations. The general $n$th moment of variance is defined as
\begin{equation} \label{eq:sigma}
s_n(z)=\sigma^2_n(R) = \frac{1}{(2\pi)^3} \int k^{2n} P(k,z) W^2(kR) d^3k,
\end{equation}
where $R$ is the smoothing scale and $P(k,z)$ is the linear matter power spectrum which depends on the wavenumber and redshift. $W(kR)$ is the Fourier transform of the window function. In this work, we use the Gaussian window function (i.e.$W(kR) = e^{-k^2R^2/2}$) through our calculations.
In the case, if the barrier is a scale-dependent quantity, which is the case when we assume an ellipsoidal collapse, The fraction of up-crossed trajectories is well fitted by Sheth and Tormen fitting function which is \citep{Sheth:1999mn}
\begin{equation} \label{eq:ST}
\nu f_{ST}(\nu) = A\left(1+(\tilde{\nu}^2)^\alpha\right) \sqrt{\dfrac{2}{\pi}} \tilde{\nu} e^{-\tilde{\nu}^2 / 2},
\end{equation}
where $\tilde{\nu}=\sqrt{q} \nu$ and $\alpha=-0.3$. Theoretical estimation for $q$, from the ellipsoidal collapse approach is, $q=1$, but by comparing with the number density of halos in simulations, we will get the result of $q=0.322$ \citep{Sheth:1999mn}. Note that the non zero $q$ could be noted as the departure from universality.\\ 
For non-Markov processes, which arise when one chooses a window function other than k-space sharp filter, it's hard to find the first up crossing probability. There is a suggestion to use the up-crossing approximation instead of first up-crossing \citep{Musso:2012qk}. This approximation works well for large masses (low variances), however for small masses when the variance of perturbations is large, this approximation breaks. For up-crossing approximation, we define another parameter $x$ which encodes the derivative of $\nu$, the height parameter, as $x=2\gamma \sigma_0 \dfrac{d \delta}{d s_0}$
where $\gamma=\sigma_1^2/\sigma_0 \sigma_2$. Counting the number of trajectories which cross a constant barrier and having the condition of $x\geq 0$ give the approximate result for a fraction of mass of the cosmic objects under consideration (e.g. halos or filaments). The final result due to \citep{Musso:2012qk} is as below
\begin{equation} \label{eq:MS}
\nu f_{MS}(\nu) = \dfrac{\langle x|\gamma,\gamma \nu \rangle_{MS}}{2\gamma \nu}\sqrt{\dfrac{2}{\pi}} \nu e^{-\nu^2 / 2},
\end{equation}
{{here $MS$ refers to Musso-Sheth proposal and
\begin{equation} \label{eq:x-x*}
\langle x|\gamma,x_* \rangle_{MS} = \int^{\infty}_0 \, dx \, x \, p_G(x-x_*;1-\gamma^2),
\end{equation} }}
 in which $p_G(x-x_*,s)$ is a Gaussian distribution function with mean of $x_*$ and variance of $s$. The Eq.~(\ref{eq:x-x*}) could be solved analytically to find
\begin{equation} \label{eq:x-x*1}
\langle x|\gamma,x_* \rangle_{MS} = \sqrt{\dfrac{1-\gamma^2}{2\pi}} \exp\left(-\epsilon^2\right)  + \dfrac{\gamma \nu}{2} \left(1+\xi\left(\epsilon \right) \right),
\end{equation}
where $\epsilon=\gamma \nu/\sqrt{2(1-\gamma^2)}$ and $\xi(x)$ is the error function. The advantage of MS approximation is that the calculations are done in one scale and it does not depend on the previous steps in the density contrast versus variance trajectories. In another word, one does not need to find the behavior of trajectories in all larger scales, which is essential in non-Markovian processes. One step further, we can improve the result by changing the $\nu$ with $\tilde{\nu}$ which have defined in Eq.(\ref{eq:ST}). This change of the variable is very similar to the ellipsoidal collapse case which fixes the deviation of MS solution from the ST in large mass scales.\\
Recently \citep{Nikakhtar:2018qqg} introduced a systematic way to go beyond MS approximation known as Hertz and Stratonovich approximation.
But in this work, we will use the idea of up-crossing to solve the structure in structure problem which will be raised in filaments.
For future work it's possible to go beyond this approximation as implemented in \citep{Nikakhtar:2018qqg}.
In the next subsection we will discuss the excursion set theory of peaks.
\subsection{Excursion Set theory of peaks}
In the previous subsection, we found the probability of collapsed over-dense regions by counting the fraction of trajectories that up crosses the linear theory critical density barrier. The main challenge in this point of view is that how we can find the first up crossing of trajectories? As we know, in the context of EST, we are not sensitive to the position of trajectories in real space, despite the fact that the formation of the objects tends to be in the extremum of density field.
In order to consider the real space distribution of the density field and its effect on the process of structure formation, we could count the number density of peaks or saddle point of the initial Gaussian density \cite{Bardeen:1985tr}. All the information is encoded in the joint probability distribution of field parameters as
\begin{equation}\label{eq:probdist}
	P(\nu,\boldsymbol{\eta},x,y,z)d\nu d^{3}\boldsymbol{\eta} dx dy dz =N|2y(y^{2}-z^{2})|e^{-Q}d \nu dx dy dz \dfrac{d^{3}\eta}{\sigma_{0}^{3}},
\end{equation}
where $\nu$ is the height parameter (i.e. $\nu = \delta_c/\sigma_0$) in this context as well, $\eta$ is the derivative of the density contrast with respect to position $\eta =\partial_i \delta$ and $(x,y,z)$ are three parameters which define the second order derivative of the density field with respect to the position $\zeta_{ij} =\partial_{ij} \delta$ as below
\begin{align} \label{eq:x,y,z}
x =& -\dfrac{\nabla^2 \delta}{\sigma_2} \\
y =& -\dfrac{\partial_x^2-\partial_z^2}{2\sigma_2}\, \delta \\
z =& - \dfrac{\partial_x^2-2\partial_y^2+\partial_z^2}{2\sigma_2}\, \delta
\end{align}
Comparing Eq.~(\ref{eq:x,y,z}) with the principal axes of $\zeta_{ij}$ using Eq.~(\ref{eq:e-p}) we find that $y/x= e_{\delta}$, $z/x= p_{\delta}$. In Eq.~(\ref{eq:probdist}) $N$ and $Q$ are
\begin{align}\label{eq:Q,N}
	N =& \dfrac{(15)^{5/2}}{32\pi^{3}}\dfrac{\sigma_{0}^{3}}{\sigma_{1}^{3}(1-\gamma^{2})^{1/2}}, \\
	2Q =& \nu^2+\dfrac{(x-x_*^2)}{(1-\gamma^2)}+15y^2+5z^2+\dfrac{3\boldsymbol{\eta}.\boldsymbol{\eta}}{\sigma_1^2}.
\end{align}
Integrating Eq.~(\ref{eq:probdist}) with appropriate constraints, for example, peak condition, we can find the number density of peaks in initial density.
\begin{equation} \label{eq:peaknumberdensity}
\mathcal{N}_{pk}(\nu) = \int_{\nu}^{\infty} d\nu \int_0^{\infty} dx  \int_{\{y,z\}\in pk} dy dz |\det(\zeta)| P(\nu,\boldsymbol{\eta}=0,x,y,z),
\end{equation}
In Eq.~(\ref{eq:peaknumberdensity}) the factor of $|\det(\zeta)|$ is necessary in transforming the result to real space. We can solve Eq.~(\ref{eq:peaknumberdensity}) analytically to find
\begin{equation} \label{eq:numxnu}
	\mathcal{N}_{pk}(\nu)
	= \dfrac{e^{-\nu^{2}/2}}{(2\pi)^{2}R_{\ast}^{3}} \int_{0}^{\infty} dx f(x) p_G(x-x_*;1-\gamma^2)
\end{equation}
where
\begin{eqnarray} \label{eq:halo}
\begin{split}
	f(x)=(x^{3}-3x) \left\lbrace \xi \left[ \left(\dfrac{5}{2}\right)^{1/2}x\right]+\xi \left[\left(\dfrac{5}{2}\right)^{1/2}\dfrac{x}{2}\right] \right\rbrace  /2\\ +\left(\dfrac{2}{5\pi}\right)^{1/2}\left[\left(\dfrac{31x^{2}}{4}+\dfrac{8}{5}\right)e^{-5x^{2}/8}+\left(\dfrac{x^{2}}{2}-\dfrac{8}{5}\right)e^{-5x^{2}/2}\right].
\end{split}
\end{eqnarray}
Note that in peak theory it is not straightforward to find mass fraction by the knowledge of the number density of peaks, because in peak theory, the scale is set and it does not change. However, we can do this naively to find the mass fraction of peaks  by the proposal of Bardeen, Bond, Kaiser and Szalay (BBKS) \citep{Bardeen:1985tr}
\begin{equation} \label{eq:BBKS}
f_{\text{BBKS}}(\nu) =e^{-\nu^{2}/2} \dfrac{V_R}{(2\pi )^{2}R_{\ast}^{3}} \int_{0}^{\infty} dx f(x) p_G(x-x_*;1-\gamma^2),
\end{equation}
where $V_R = (2\pi R^2) ^{3/2}$ is the volume of smoothing region with scale $R$ and $R_{\ast} = \sqrt{3}\sigma_1/ \sigma_2$. The BBKS is successful in counting the number of peaks but hardly deal with the "cloud in cloud" problem, which its solution must be analogue to the first up crossing condition in EST. The reason is that in peak condition, it is only one scale which is involved and by changing the scale the peak even could be disappeared. However, by reducing the condition of first up crossing to the condition of just up crossing, we could overcome this problem.\\
Comparing  Eq.~(\ref{eq:BBKS}) with Eq.(\ref{eq:MS}) leads us to understand the  differences. The term $f(x)$ in BBKS is responsible for peak condition and extra $x$ in MS is responsible for up crossing condition. Here we must mention that $x$ in Eq.~(\ref{eq:BBKS}) is not the same as $x$ in Eq.~(\ref{eq:MS}) except when the window function is Gaussian which is the case of our work. So by working with Eq.~(\ref{eq:BBKS}) and adding the extra $x$ to it, we can add the up crossing condition to BBKS as proposed by \citep{Paranjape:2012ks}
\begin{equation} \label{eq:ESP}
f_{\text{ESP}}(\nu) =e^{-\nu^{2}/2} \dfrac{V_R}{(2\pi )^{2}R_{\ast}^{3}} \int_{0}^{\infty} dx\, x  f(x) p_G(x-x_*;1-\gamma^2),
\end{equation}
hereafter the "ESP" refers to excursion set of peaks. In the next subsection, we will introduce a new concept named as excursion set of saddle points as a framework to study the distribution of filaments.

\subsection{Excursion set of saddle points}
In the context of the ellipsoidal collapse model, filaments are considered as objects which have two collapsed axes. This assumption is made without any reference to the place of filaments in the cosmic web. As we mentioned and discussed in the introduction the filaments tend to form in the saddle points of matter density or saddle points of potential field in order to take the effect of  environment in to account. In this work, we consider the first criteria as the the definition of the filaments. However, we should mention that whenever the perturbations are very near to homogeneous and isotropic case, the saddle points of potential occur near to the saddle points of matter density field and vise versa, when the Poisson equation relates density contrast to gravitational potential.
So the difference might be neglected in this case.\\
In order to count the number of the saddle points in the initial density field, we start with an equation very similar to the one introduced in Eq.~(\ref{eq:peaknumberdensity}). Now the peak condition must be changed to the condition which points out the saddle points with two positive and one negative eigenvalues $\{\lambda_1^{\delta},\lambda_2^{\delta}\geq 0,\lambda_3^{\delta}\leq 0\}$ accordingly, we will get
\begin{equation} \label{eq:saddlenumberdensity}
\mathcal{N}_{\text{sd}}(\nu) = \int_{\nu}^{\infty} d\nu \int_0^{\infty} dx  \int_{\{y,z\}\in sd} dy dz |\det(\zeta)| P(\nu,\boldsymbol{\eta}=0,x,y,z),
\end{equation}
where "sd" refers to saddle point condition. Again the integral can be simplified as
\begin{equation} \label{eq:numxnusd}
	\mathcal{N}_{sd}(\nu)
	= \dfrac{e^{-\nu^{2}/2}}{(2\pi)^{2}R_{\ast}^{3}} \int_{0}^{\infty} dx h(x) p_G(x-x_*;1-\gamma^2),
\end{equation}
where $h(x)$ is the factor which is responsible for the saddle point condition. There is an analytical solution for $h(x)$ as
\begin{equation}\label{eq:saddle}
h(x)=\frac{e^{-\frac{5 x^2}{8}} \left(32+155 x^2\right)}{10 \sqrt{10 \pi }}-\frac{1}{2} x \left(-3+x^2\right) erfc \left[\frac{1}{2} \sqrt{\frac{5}{2}}
x \right],
\end{equation}
where $erfc$ is the complementary error function. We can count the number of the collapsed saddle point by applying the condition of $v\geq \delta_c/\sigma_0$ to Eq.(\ref{eq:numxnusd}). However similar to the case of halos, there is the cloud in cloud problem, which literally can be named as the filament in filament problem. In this point, one can use the resemblance of the process to the halo case and get rid of this problem by using up crossing approximation. So we use the idea introduced in Eq.~(\ref{eq:ESP}) to write
\begin{equation} \label{eq:ESS}
f_{\text{ESS}}(\nu) =e^{-\nu^{2}/2}\dfrac{V_R}{(2\pi )^{2}R_{\ast}^{3}} \int_{0}^{\infty} dx\, x  h(x) p_G(x-x_*;1-\gamma^2),
\end{equation}
here "ESS" refers to excursion set of saddle points and $h(x)$ is defined in Eq.~(\ref{eq:saddle}).
\begin{figure}
	\includegraphics[width=\columnwidth]{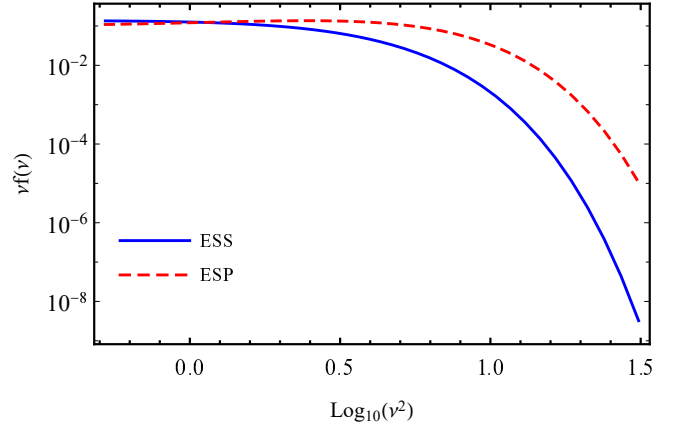}
    \caption{ The fraction of up-crossing $\nu f(\nu)$ for excursion set of peaks (red dashed line) and the excursion set of saddle points (solid blue line) is plotted versus the square of height parameter in logarithmic scale  $\log_{10} \nu^2$  }
    \label{fig:saddlepoint}
\end{figure}
In Fig.~\ref{fig:saddlepoint}, we plot the fraction of up-crossing in the context of excursion set theory of peaks (which then must be interpreted as the number density of halos) and excursion set of saddle points in terms of height parameter (which is related to the filaments). The $\nu f(\nu)$ is a key quantity, from which we can find the number density of filaments accordingly. Now the number density of filament $n(M)$ could also be calculated as
\begin{equation} \label{eq:numberdensity}
n(M)\, dM = \dfrac{\bar{\rho}}{M} f(s) ds \left| \dfrac{ds}{dM} \right| \,dM,
\end{equation}
where $f(s)$ is up-crossing fraction in the variance interval of $s$ and $s+ds$.
\begin{figure}
	\includegraphics[width=\columnwidth]{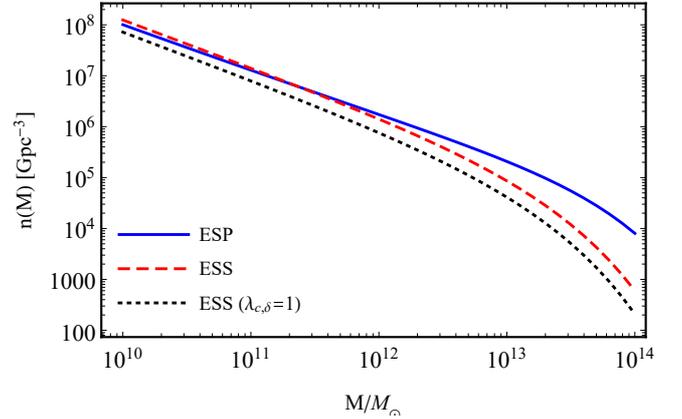}
    \caption{The number density of filaments (solid blue line) in the context of excursion set of saddle points is plotted versus mass. For comparison, the number density of halos in the context of excursion set of peaks is also plotted (red dashed line). {The black dotted line is the the number density with the constraint of filament's extended condition and $\lambda_{c,\delta} = 1$.}}
    \label{fig:massnumber}
\end{figure}
In Fig.~\ref{fig:massnumber} we plot the number density of filaments in terms of mass in the context of excursion set of saddle points. For comparison in the same plot, we show the number density of halos obtained from excursion set of peaks accordingly. The plot shows that in larger masses with considering a specific collapsed mass the number density of filaments is smaller than the number density of halos. This is a reasonable result because in the context of ellipsoidal collapse and peak/saddle theory the large masses tends to become peaks (halos) instead of saddle points. {In this plot we also show the filament's extended condition case ($\lambda_{c,\delta} = 1$) with dotted black line. The result agrees with our intuition that making constraints tighter will reduce teh number density of filamnets.} \\
We could use Eq.(\ref{eq:filamentlenght}) to find the length number density of filaments in this approach.
\begin{figure}
	\includegraphics[width=\columnwidth]{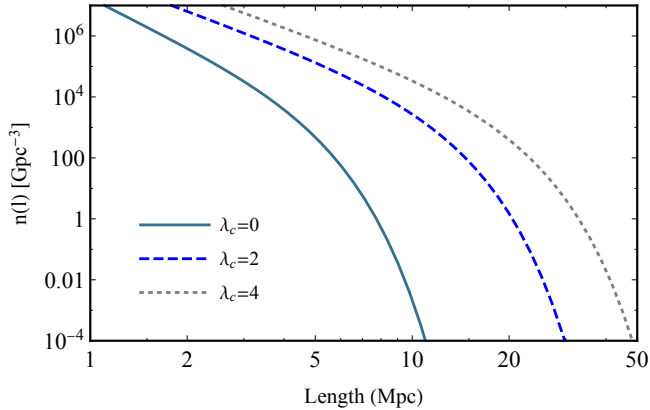}
    \caption{{The number density of filaments is plotted versus length for different cases. The solid green line for $\lambda_{c,\phi} = 0$, Blue dashed line for $\lambda_{c,\phi} = 2$ and black dotted line for $\lambda_{c,\phi} = 4$.}}
    \label{fig:lengthnumber}
\end{figure}
{In Fig.(\ref{fig:lengthnumber}), we plot the number density of filaments versus their length. This is obtained by exchanging the mass in the Eq.(\ref{eq:numberdensity}) with length using Eq.(\ref{eq:filamentlenght}) in which we replaced $e_{\phi}$ and $p_{\phi}$ by their averaged value with appropriate conditions( The solid green line is for $\lambda_{c,\phi} = 0$, blue dashed line corresponds to $\lambda_{c,\phi} = 2$  and $\lambda_{c,\phi} = 4$  is shown by black dotted line). As we anticipate the number density of filaments is a decreasing function of length. Filaments with large lengths correspond to over-dense regions with larger mass. We know that for large ellipsoid over-dense regions, there is a tendency to be virialized in halos. As it is obvious in this figure increasing the amount of $\lambda_{c,\phi}$ causes the number density of filament being increased for a defined length. This is because by increasing the $\lambda_{c,\phi}$ we assign larger filaments length to a specific mass. This means that for larger $\lambda_{c,\phi}$, a chosen length refers to  a smaller mass. Finally, the number density is a monotonic increasing function with respect to $\lambda_{c,\phi}$ as the number density is a monotonic decreasing function of mass}.\\
This plot can be used as a crucial proposal for future upcoming large scale structure surveys, where with observing large number of galaxies, the statistics of filaments become considerably large for comparison with theoretical models.
{Once again, we should emphasis here that our proposal is a local treatment of the filaments as the saddle point of density field, where the condition on $\lambda_{c,\phi}$ is an approximation to have two massive dark matter halos in the endpoints. This approximation which is refereed to filaments extended condition change the mass-length relation and the number density of filaments. Accordingly a non-local treatment of filaments is essential. }

In App.\ref{app:moving} we will study the number density of filaments with considering the moving barrier for up-crossing. \\
\section{Conclusions and Future remarks}
\label{Sec:conclusion}
The future large scale structure surveys will open up a new horizon to study of the cosmic web in detail. Due to simulations and observations, it seems obvious that beside the dark matter halos, the dark filaments can be considered as promising large cosmic scale structures which can be used to study the formation and evolution of cosmic skeleton. On the other hand, the measurable characteristics of filaments can be used as a probe to test cosmological models. In this work, we study the filaments in the context of ellipsoidal collapse. Using the conditions on the eigenvalues of Hessian matrix of density field we define the regions in ellipticity-prolateness plane for which we can make halos, filaments, and sheets. {In this direction we introduce "filament's extended condition" which improve our result especially in comparing mass-length relation of our model and simulation results. This extra condition helps us to extend our local study of filaments as saddle points to the one which consider the fact that filaments are ended with massive dark matter halo nodes.} \\
We also consider that the critical density of filaments formation depends on prolateness and has a negligible dependence on ellipticity. Then using the concept of ellipsoidal collapse we obtain the length of filaments versus their mass and a fitting function for length-mass relation in flat $\Lambda$CDM cosmology. The non-linear density contrast of filaments are obtained, where we show that this non-linear density contrast is almost mass independent and it can be used as a proxy to find the filaments in the simulation. In this work, we also address the question of the number density of filaments. For this task, we define the formalism of the excursion set of saddle points instead of excursion set of peaks in order to count the saddle points and also we use the idea of up-crossing approximation to solve the known obstacle of structure in structure problem. Accordingly, we find the number density of filaments in terms of mass and length. This study is crucial as it is an analytical suggestion to study the filaments in non-linear structure formation framework to address cosmological questions and to understand the physics of filament formation. We should note that this result is in the context of ellipsoidal collapse {and in this context the structure under consideration, initially, is approximately spherical. This assumption may contradicts the fact that filaments have elongated thin structure at very early stages of their formation which evolve to thicker structures \citep{Bond:1995yt}. In this direction we point out the role of smoothing window function which make thin elongated objects more spherical. However, this concern may be a real caveat of our model which must be improved in future works.}
As a future remark, we suggest that sophisticated filament finder procedure can be used in simulation to compare the predictions of analytical approach we developed here.

\section*{Acknowledgements}

We would like to thank Ali Akbar Abolhasani, Ehsan Ebrahimian, Marcello Musso, Farnik Nikakhtar, Sohrab Rahvar, and Ravi K. Sheth for useful comments and fruitful discussion. We also want to thank the anonymous referee for his/her insightful comments.  MA and SB would like to thank the Abdus Salam International Center of Theoretical Physics (ICTP) for a  kind hospitality, which the initial ideas of this work is formed there. MA and SB thank the school of Astronomy of Institute for Research in Fundamental Sciences (IPM) for hospitality during this work. SB is partially supported by Abdus Salam International Center of Theoretical Physics (ICTP) under the junior associateship  scheme during this work. This research is supported by Sharif University of Technology Office of Vice President for Research under Grant No. G960202.




\appendix

\section{Filament length in ellipsoidal collapse} \label{app:length}
In the ellipsoidal collapse, the main idea is to consider the evolution of a homogeneous ellipsoid. The differential equation governing the dynamics of the principal axes of the ellipsoid is as below \citep{peebles1980,BondMayers1996}
\begin{equation} \label{Eq: ellipsoidal collapse}
\dfrac{d^2 X_j}{dt^2} + \dfrac{2 \dot{a}}{a} \dfrac{dX_j}{dt} = -4\pi G \bar{\rho}_m X_j \left[ \dfrac{1}{2} \alpha_j \delta + D(t) (\lambda_j - \dfrac{1}{3}\delta_i) \right],
\end{equation}
In which
\begin{equation} \label{Eq: alphai}
\alpha_i = X_1 X_2 X_3 \int \dfrac{dy}{(X_i^2+y)f^{1/2}(y)},
\end{equation}
and
\begin{equation} \label{Eq: f}
f(y) = (X_1^2 + y)(X_2^2+y)(X_3^2+y).
\end{equation}
Here $X_j$ is the normalized length of axes of filament and the comoving length could be defined
\begin{equation} \label{Eq: r0}
a_j = \left(\beta_W \dfrac{GM}{H_0^2 \Omega_m} \right)^{1/3} X_j,
\end{equation}
where $\beta_W$ is defined in the main text and depends to window function.
We must mention that Eq.~(\ref{Eq: ellipsoidal collapse}) is well working in both linear and non linear regime and also for spherical case $\lambda_1 = \lambda_2 = \lambda_3$.
\begin{figure}
\centering
\includegraphics[width=0.45\textwidth]{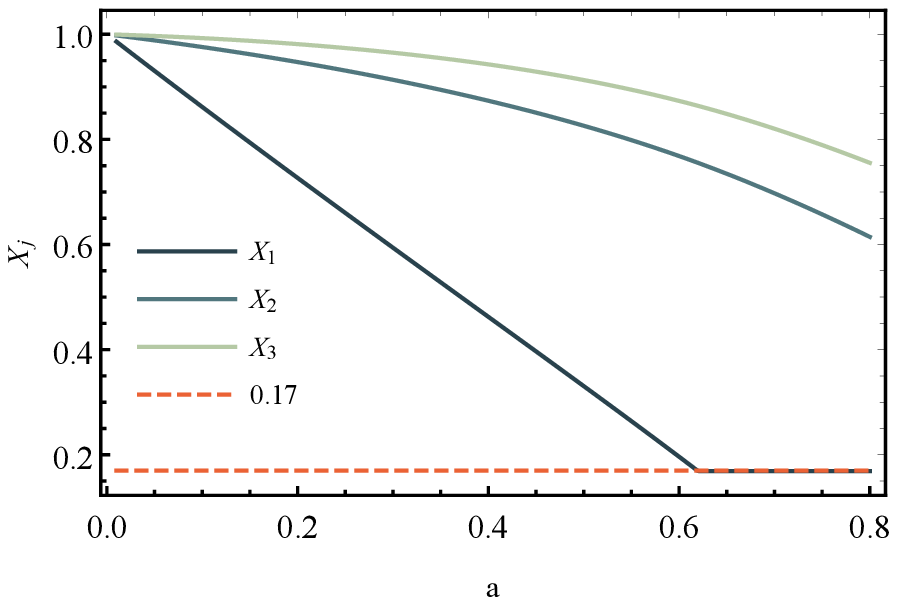}
\caption{The length of the three axis of ellipsoid over-dense region is plotted in terms of scale factor with $e= 0.4$ and $p=0.3$ (in halo regime). The red dashed red line corresponds to the virial length of a halo.}
\label{FIG: ellipsoidal collapse}
\end{figure}
\begin{figure}
\centering
\includegraphics[width=0.45\textwidth]{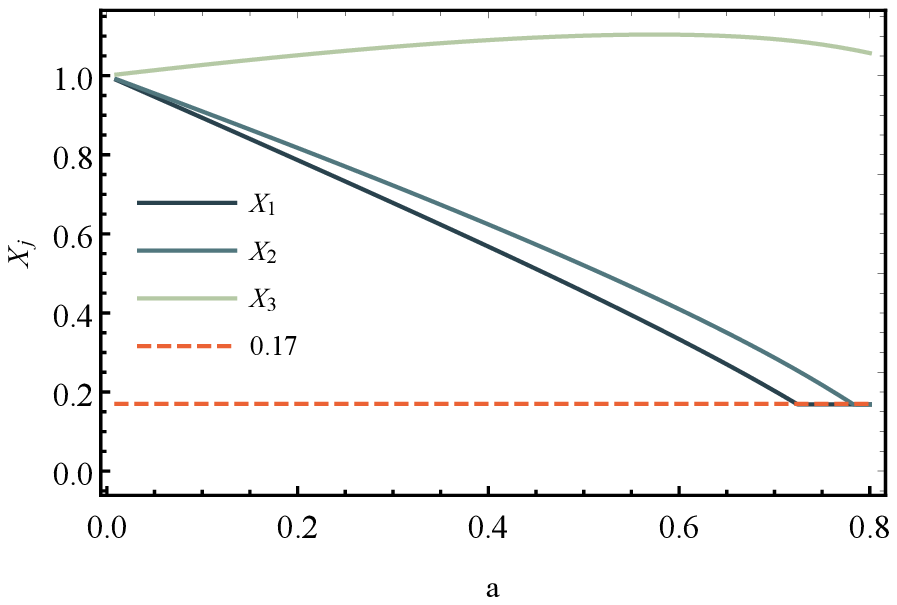}
\caption{The length of the three axis of ellipsoid in terms scale factor $e= 0.4$ and $p=-0.3$ (in filament range). The dashed  red line corresponds to the virial length of a halo.}
\label{FIG: ellipsoidal collapse1}
\end{figure}
One can solve Eq.~(\ref{Eq: ellipsoidal collapse}) numerically having initial condition in hand for $\{\lambda_1, \lambda_2, \lambda_3\}$ or interchangeably  with the initial condition on density contrast, ellipticity and prolaticity $\{\delta_i, e,p\}$ which is defined in Eq.~(\ref{eq:e-p}). Note that here we consider the ordering of the eigenvalues $\lambda_1 \geq \lambda_2 \geq \lambda_3$. The Fig~\ref{FIG: ellipsoidal collapse} shows the numerical solution of Eq.~(\ref{Eq: ellipsoidal collapse}) for a special initial condition by considering a viral length for the axes as $X_v = 0.17$ \citep{BondMayers1996,Shen:2005wd}. In Fig~\ref{Eq: ellipsoidal collapse}, we see the behavior of principal axes, which shows that how the ellipsoid forms a sheet and then a filament and finally a halo. In this figure the condition on the $e_{\phi}$ and $p_{\phi}$ lies in the halo range of Fig~\ref{fig:e,p}. In Fig~\ref{FIG: ellipsoidal collapse1} evolution of the ellipsoid is calculated for the case when $e_{\phi}$ and $p_{\phi}$ are in the filament range.\\
The Eq.~(\ref{Eq: ellipsoidal collapse}) can be approximated by  \citep{White1979}
\begin{equation} \label{Eq: White app}
X_j(t) = X_j(t_i) (1 - D(t) \lambda_j) - X_h(t_i) \left[1 - \dfrac{D(t) \delta_i}{3} - \dfrac{a_e(t)}{a(t)} \right],
\end{equation}
 where $X_h(t) =  3/\sum X_j(t)$ is a good approximation for $\alpha_j = 2 X_h/3X_j $ and $a_e(t)$ is scale factor of a universe with initial matter density of $\bar{\rho}_m(t_i)(1+\delta_i)$. First term in right hand side of Eq.~(\ref{Eq: White app}) is Zel'dovich term and depends  on the direction $j$ under consideration but the second term is the same for the three directions. Accordingly the second term will be eliminated when the diffrence of the axes $X_i - X_j$ is considered. Note that this implies Zeldovich approximation be well enough for calculating $X_i - X_j$ as
\begin{equation} \label{Eq: diff axes}
X_3(t_f) - X_2(t_f) = D(t)(\lambda_2-\lambda_3) = D(t) \delta_i (e-p),
\end{equation}
and as the linear density contrast is $\delta_f^L=D(t)\delta_i$ then
\begin{equation} \label{Eq: X=e}
X_3(t_f) = X_2(t_f) + \delta_f^L (e-p).
\end{equation}
In this context, the filament is considered as a cosmic structure which is collapsed in two dimensions and have an uncollapsed axes. It means $X_1(t_f) \leq X_v$, $X_2(t_f) = X_v$ and $X_3(t_f) \geq X_v$, Where $X_3$ is the uncollapsed principle axis of ellipsoid ( this is the filament length) and $t_f$  is the time of filament formation. So $X_2(t_f) = X_v$ and the density contrast of the filament becomes  $\delta_f = \delta_c(1+\alpha p_{\phi})$ Eq.~(\ref{eq:delta_c}). Note that the formation time of filament can be obtained from  $\delta_f = D(t_f) \delta_i$ \cite{Shen:2005wd}. Solving Eq.~(\ref{Eq: ellipsoidal collapse}) numerically or by considering Eq.~(\ref{Eq: White app}) one can find that $\delta_f$ is independent of $e$ and it's dependence over $p$ is like Eq.~(\ref{eq:delta_c}) and this is satisfied in the context of ellipsoidal collapse framework. \\
Note that the comoving length of the filament is $a_3 = r_0 X_3$, in which $r_0$ is defined in Eq.~(\ref{Eq: r0}). So Eq~(\ref{Eq: X=e}) easily read to the length of the filament,
\begin{equation}\label{Eq: filament lenght}
l(M,e,p) =  \left(\sqrt{\dfrac{8}{9\pi}}\dfrac{GM}{H_0^2 \Omega_m} \right)^{1/3} [X_v + \delta_f (e-p)],
\end{equation}
It is a reasonable approximation to change $\delta_f$ with $\delta_c$ in Eq.(\ref{Eq: filament lenght}), so the result will be the same as Eq.(\ref{eq:filamentlenght}).

\section{Number density of filaments with moving barrier} \label{app:moving}
In Sec.(\ref{Sec:ESS}), we have considered a constant barrier (critical density) in our calculation to obtain the number density of filaments. The reason for choosing a constant barrier is that when we average over all regions in $e-p$ space we will have $\langle p\rangle=0$ and considering the Eq.~(\ref{eq:delta_c}) implies a constant barrier.  In this appendix, we will come back to the case which the barrier is not constant due to the fact that averaging is done over the filament region in  $e-p$ space. This approach is necessary, if we have some constraint on the potential density field (as we consider throughout of this work). For the moving barrier the Eq.~(\ref{eq:ESS}) must be changed with the equation below \citep{Paranjape:2012ks}
\begin{align} \label{eq:ESP-Moving}
f_{MB}(\nu) =e^{-\nu^{2}_f/2} \dfrac{(2\pi)^{2}R^{3}}{(2\pi)^{2}R_{\ast}^{3}}& \int_{2\gamma \sigma_0 \langle \delta'_f \rangle}^{\infty} dx\, \\  \nonumber
&(x-2\gamma \sigma_0 \langle\delta'_f\rangle)  f(x) p_G(x-x_{*f};1-\gamma^2),
\end{align}
where "MB" indicates the moving barrier case, $\langle \delta_f \rangle$ is averaged critical density of Eq.~(\ref{eq:delta_c}). $\nu_f$ and  $x_{*f}$ are corresponding variables.
\begin{figure}
	\includegraphics[width=\columnwidth]{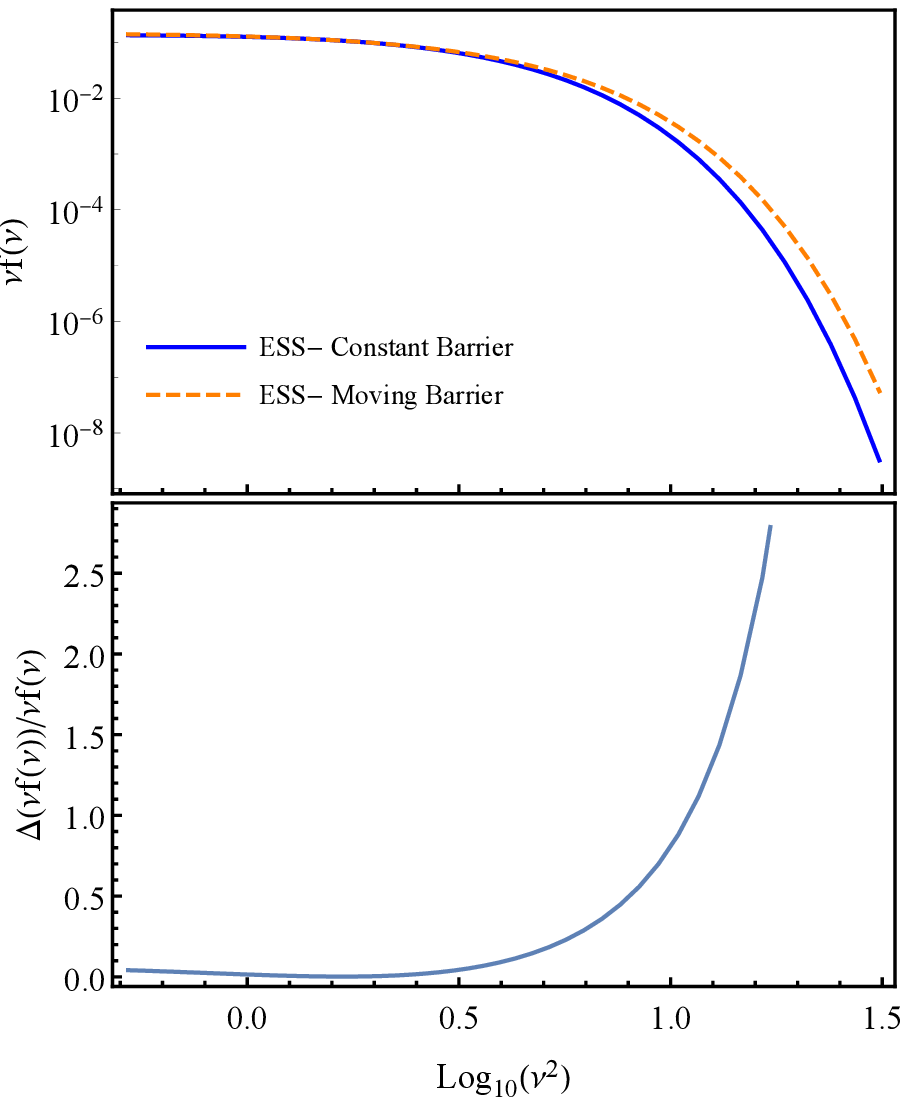}
    \caption{{In upper figure, number density of filaments is plotted versus length with constant barrier condition (blue solid line) and with the moving barrier condition (dashed orange line). The relative difference is plotted in the bottom figure.}}
    \label{fig:fraction-moving}
\end{figure}
{In upper plot of Fig~(\ref{fig:fraction-moving}) the mass fraction of the filaments is plotted for two cases of the moving and constant barrier. In a logarithmic scale, the moving condition increases the mass fraction in large $\nu$ and masses. The relative difference is plotted in bottom plot of Fig~(\ref{fig:fraction-moving}). For this figure, we used the filament constraint of Fig.(\ref{fig:e,p}).}\\
\bsp	
\label{lastpage}
\end{document}